\documentclass[epj]{webofc}
\usepackage[varg]{txfonts}   
\usepackage{graphicx,color} 
\usepackage{bm}       
\usepackage{amsmath}  
\usepackage{amssymb}
\usepackage{comment}
\usepackage{wrapfig}

\newcommand{\blf}[1]{\bf  {\tilde #1}}
\newcommand{\bq}{\begin{eqnarray}}
\newcommand{\eq}{\end{eqnarray}}

\woctitle{21st International Conference on Few-Body Problems in Physics}
\begin{document}
\title{The $^3$He spectral function in
light-front dynamics
}
\author{Matteo
Rinaldi\inst{1}\fnsep\thanks{\email{matteo.rinaldi@pg.infn.it} } \and
        Alessio Del
Dotto\inst{2}\fnsep \and
    Leonid
Kaptari\inst{3}\fnsep \and Emanuele Pace \inst{4} \and Giovanni Salm\`e
\inst{5} \and Sergio Scopetta \inst{1}
}

\institute{Dipartimento di Fisica e Geologia, Universit\`a di Perugia and
INFN, Sezione di Perugia, Italy
\and
           Dipartimento di Fisica, Universit\`a di Roma Tre and INFN, Sezione 
Roma
3, Italy 
\and
           Bogoliubov Lab. Theor. Phys., 141980, JINR, Dubna, Russia
\and
Dipartimento di Fisica, Universit\`a di Roma ``Tor Vergata'' and INFN, Sezione 
Roma
2, Italy
\and
INFN Sezione di Roma
          }

\abstract{
A distorted spin-dependent spectral function for $^3$He is considered
for the extraction of the transverse-momentum 
dependent parton distributions in the neutron 
from semi-inclusive
deep inelastic electron scattering off polarized $^3$He at finite momentum
transfers, 
where final state
interactions 
are taken into account.
The generalization of the
analysis to a Poincaré covariant framework within the light-front dynamics is
outlined. 
}
\maketitle
%

In the next few years, several experiments involving
$^3$He nuclear targets will be performed at JLab, 
with the aim of extracting information on the parton structure
of the neutron.
In particular, the 
three-dimensional neutron structure,
described in terms of quark transverse momentum 
dependent parton distributions (TMDs) 
\cite{uno}, will be
probed through spin-dependent
semi-inclusive deep inelastic scattering (SIDIS) processes off
$^3$He, where
a high-energy pion is detected in coincidence with the scattered electron
\cite{cinque}.
Hence, to clarify the flavour dependence
of the TMDs, high precision 
experiments, involving both protons and neutrons,
are needed.
To obtain a reliable information from $^3$He SIDIS
one has to take into account: 
i) the nuclear structure
of  $^3$He, ii) the interaction in the final state (FSI) between the observed
pion and the remnant
debris, and iii) the relativistic effects. Our
efforts about the above issues are summarized here.
 
An initial study of this kind was reported in 
Ref. \cite{nove},
where the plane wave impulse approximation (PWIA)
was adopted to describe
the reaction mechanism, treating the $^3$He structure 
non-relativistically, using the AV18 interaction \cite{a18}.
In facts,
a polarized $^3$He is an ideal target 
to study the {{neutron}}, since at a 90\% level a polarized $^3$He 
is equivalent to
a free polarized neutron \cite{ciofi}.
In Ref. \cite{nove} it was found, using a realistic
$^3$He
spin-dependent spectral function, ${{ P^{\tau}_{\sigma,\sigma{\prime}} (\vec p,
E)}}$, with  
$\vec p$ the nucleon momentum in the laboratory frame and $E$ 
the removal energy \cite{Kiev},
that the formula
\begin{equation}
{{A_n }}\simeq \left 
( {A^{exp}_3} - 2 
{p_p} f_p ~
{{A^{exp}_p}} \right )/ (p_n f_n)
\end{equation}
can be safely adopted to extract the neutron 
Sivers and Collins 
asymmetries \cite{sivers,collins}
from $^3$He and proton data.
This formula
has been already used
by experimental collaborations \cite{Qian,Allada}.
Nuclear effects are hidden
in the proton and  neutron {{"effective polarizations"}} (EPs), 
$ p_{p(n)}$, and 
in the dilution factors, $f_{p(n)}$, properly defined
\cite{nove}. With the AV18 interaction, 
one obtains
$ {{p_p}} = -0.023  $, 
${{p_n}}= 0.878 $ \cite{nove}.
\begin{wrapfigure}{l}{0.4\textwidth} 
 \includegraphics[width=6cm]{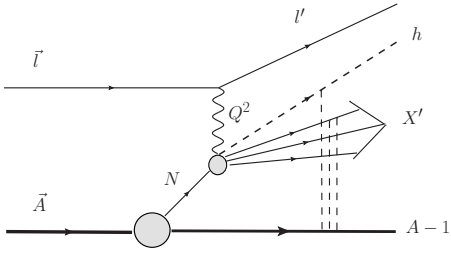}
 \caption{Interaction between the $(A-1)$ spectator system and the debris
produced by the
absorption of a virtual photon by a nucleon in the nucleus.}
\end{wrapfigure}

We are now taking into account
FSI between the observed pion and the remnants.
In a recent paper \cite{undici}, the distorted spin 
dependent spectral function
of $^3$He has been introduced,
and evaluated when the final $A-1$ system is a deuteron state.
The evaluation considering the most general final state
is in progress \cite{19}.
A few preliminary results are reported here.
In SIDIS off $^3$He, the relative energy between the spectator
$(A-1)$ system
and the system composed by the detected pion and the remnant debris 
(see Fig. 1)
is a few GeV. 
FSI can be therefore treated through a generalized eikonal
approximation (GEA) \cite{undici}. The GEA was already successfully applied
to unpolarized SIDIS in Ref. \cite{Kaptari}.
In GEA, FSI effects are due to the propagation of the debris,
formed after the $\gamma^*$ absorption by a target quark, and the subsequent 
hadronization, both
of them influenced by the presence of a fully interacting $(A-1)$ 
spectator system (see Fig. 1).  
The key quantity to describe FSI effects is the {\em{distorted}} 
spin-dependent spectral function, whose
relevant part in the evaluation of single spin asymmetries (SSAs) is:
${P}_{}^{PWIA({{FSI}})}=
{\cal O}^{IA({{FSI}})}_{\frac12 \frac12}-
{\cal O}^{IA({{FSI}})}_{-\frac12 -\frac12}$
with:
\vskip -3mm
\begin{eqnarray}
 \hskip -5mm {\cal O}^
{IA}_{\lambda\lambda'}(\vec p,E) ~
= \sum \! \!\! \!\! \!\! \!\int{~d\epsilon^*_{A-1}} ~
\rho\left(\epsilon^*_{A-1}\right)~
\langle   
S_A, 
{{\bf P_A}}
| 
\Phi_{\epsilon^*_{A-1}},\lambda',\vec p \rangle 
 \langle  
\Phi_{\epsilon^*_{A-1}},
\lambda,\vec p|  S_A,{{\bf P_A}}\rangle ~
\delta\left( E- B_A-\epsilon^*_{A-1}\right) 
\label{overlap} 
\end{eqnarray}
and
\begin{eqnarray}
\hskip -5mm {\cal O}^
{{FSI}}_{\lambda\lambda'}(\vec p,E) 
&=& \sum \! \!\! \!\! \!\! \!\int{~d\epsilon^*_{A-1}} ~
\rho\left(\epsilon^*_{A-1}\right)~
\langle   
S_A, 
{{\bf P_A}}
| ({{\hat S_{Gl}}})
\{\Phi_{\epsilon^*_{A-1}},\lambda',\vec p\} \rangle 
 \langle  
({{\hat S_{Gl}}})
\{\Phi_{\epsilon^*_{A-1}},
\lambda,\vec p\}|  S_A,{{\bf P_A}}\rangle ~
\\
\nonumber
&\times&\delta\left( E- B_A-\epsilon^*_{A-1}\right) ~,
\label{overlapfsi} 
\end{eqnarray}
where $\rho\left(\epsilon^*_{A-1}\right)$ is the density of the
$(A-1)$-system
states with intrinsic energy 
$\epsilon^*_{A-1}$, and $|  S_A,{{\bf P_A}}\rangle$
is the
ground state of the $A$-nucleons nucleus with polarization $S_A$.
${\hat S_{Gl}}$ is the {{Glauber}} operator:
\begin{eqnarray}
\, \, {{\hat S_{Gl}}} 
({\bf r}_1,{\bf r}_2,{\bf r}_3)=
\prod_{i=2,3}\bigl[1-\theta(z_i-z_1)
{{\Gamma}}({\bf b}_1-{\bf b}_i,{ z}_1-{z}_i)
\bigr]
\end{eqnarray}
and $\Gamma({{\bf b}_{1i}},z_{1i})$ the generalized {{profile function}}:
\begin{equation}
\quad 
{{\Gamma({{\bf b}_{1i}},z_{1i})}}\, =
\,\frac{(1-i\,\alpha)\,\,
{{\sigma_{eff}(z_{1i})}}} {4\,\pi\,b_0^2}\,\exp 
\left[-\frac{{\bf b}_{1i}^{2}}{2\,b_0^2}\right]~,
\label{profile}
\end{equation}
where ${\bf r}_{1i}=\{{\bf b}_{1i}, {\bf z}_{1i}\}$
with ${\bf z}_{1i} ={\bf z}_{1}-{\bf
z}_{i}$ and ${\bf b}_{1i}={\bf b}_{1}-{\bf b}_{i}$.
The models for the profile function, $\Gamma({{\bf b}_{1i}},z_{1i})$, and for
the 
effective cross section, $\sigma_{eff}(z_{1i})$, as well as the values of the
parameters 
$\alpha$ and $b_0$, are the ones 
used in Ref. \cite{Kaptari} to successfully describe deuteron 
JLab data \cite{KKW}.
$\sigma_{eff}(z_{1i})$ is the cross section for the
interaction of the formed hadrons,
$\alpha$ is the ratio of the real to the
imaginary part
of the forward scattering amplitude, $b_0$ is the slope
parameter
(cf Ref. \cite{CKK} for details).
\begin{figure}[t]
\includegraphics[width=12cm]{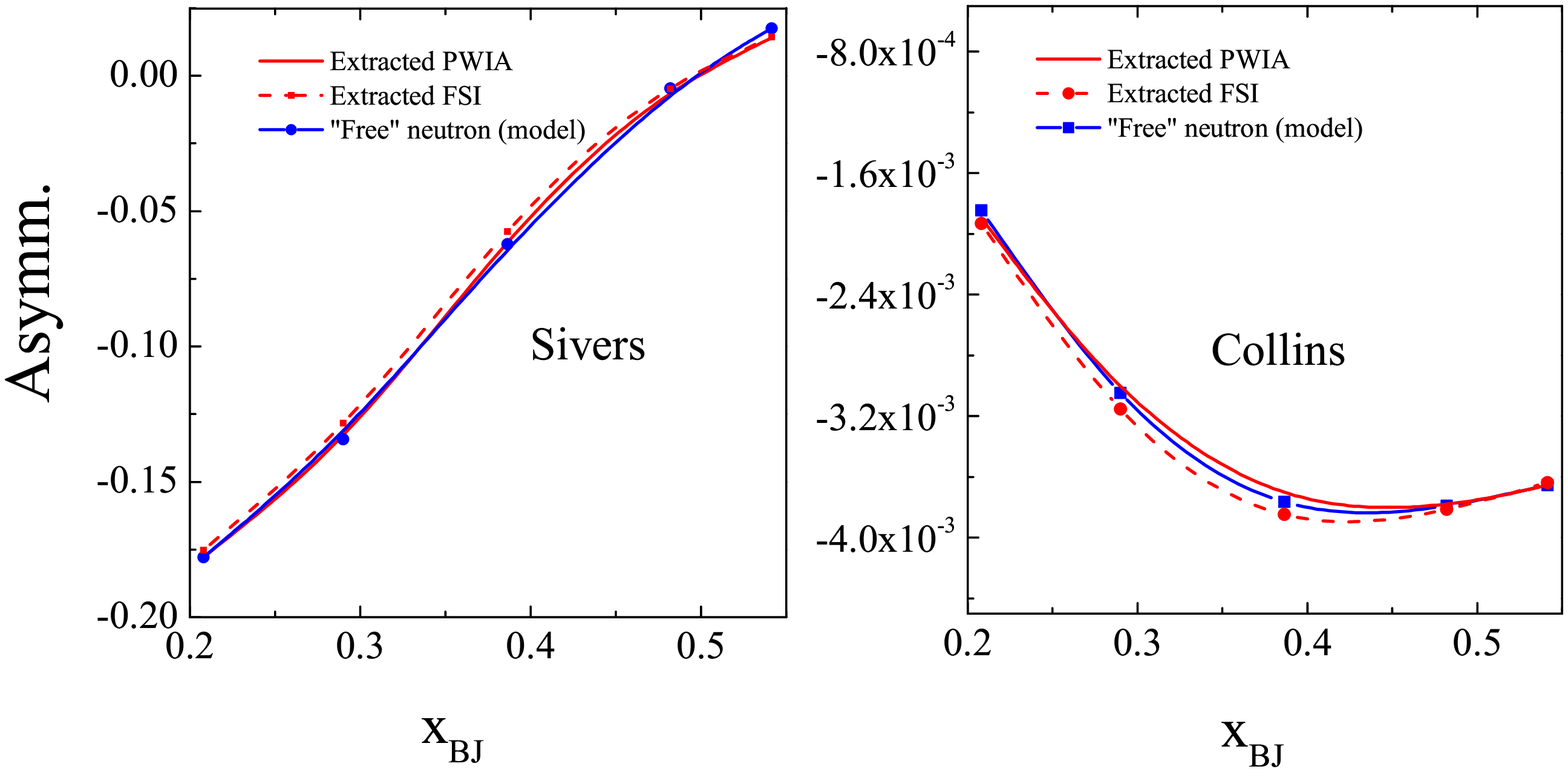}
\caption{
Effectiveness of the extraction procedure, Eq. (7), for the Sivers (left panel)
and Collins (right panel) asymmetries, with and without FSI,
in the actual kinematics of JLab \cite{cinque}.
Preliminary results to appear in \cite{19}.
}
\end{figure}
It occurs that ${P}^{PWIA}$
and 
${P}^{FSI}$
can be very different, but the relevant observables for the SSAs
involve integrals, dominated
by the low momentum region, where the  FSI effects are
minimized and the spectral function is large \cite{Kaptari}.
As a consequence, the EPs change from the $PWIA$ values, i.e.
$ {{p_p}} =
-0.023  $ and
${{p_n}}= 0.878 $,  
to ${{p_p^{FSI}}} = -0.026$, ${{p_n^{FSI}}}= 0.76 $, 
where 
\begin{equation}
  {p^{FSI}_{p(n)}} = 
  \int d\epsilon_{S} \int d{\vec p} ~ 
{ {Tr [{\bf{S}}_{He} * {\bf{\sigma}} ~
  {P}^{p(n)}_{FSI}(\vec p, \epsilon_S,S_{He}) ]
}}~. 
\label{polFSI}
\end{equation}
Then $p_{p(n)}$ with and without the FSI differ
by 10-15\% . 
Actually, one has also to consider the 
effect of the FSI on dilution factors. We
have found,
in a wide range of kinematics typical for the experiments at JLAB
\cite{cinque}, 
that the product
of polarizations and dilution factors changes very little \cite{Deldotto}.
Indeed the effects of FSI on the dilution factors and on the
EPs are found to compensate each other
to a large extent and the 
{{usual extraction}} seems to be safe, viz. :
\begin{eqnarray}
  {{A_n }}\simeq 
  \left ( {A^{exp}_3} - 2 {p_p^{FSI} f_p^{FSI} {{A^{exp}_p}}} \right )
/ (p_n^{FSI} f_n^{FSI})
~ \simeq
  \left ( {A^{exp}_3} - 2 {p_p f_p ~ {{A^{exp}_p}}} \right )
/ (p_n f_n)~, 
~ \quad   
\vspace{-0.1cm}
\end{eqnarray}
as one can clearly see in Fig. 2.

We now address the relativistic description of the 
parton structure of $^3$He, within a novel approach which, 
due to the high energies involved 
in the forthcoming experiments \cite{cinque}, is very timely \cite{14,PSS}.
To study relativistic effects, we adopt the  Light-Front (LF) form of the
Relativistic Hamiltonian Dynamics (RHD)  for an interacting system 
with a fixed number of on-mass-shell constituents. 
The Bakamijan-Thomas 
construction of the Poincar\'e generators leads to a 
fully Poincar\'e covariant RHD \cite{14,PSS}.
The  Light-Front  form of RHD has a {{subgroup}} structure of
the {{LF boosts}}, which
allows a separation of the {{intrinsic motion}} from
the global one,  very important for the description of DIS, SIDIS and 
deeply virtual Compton scattering processes. 
We have not taken into account a possible violation of cluster separability
in our approach \cite{kister}.
The key quantity to consider in SIDIS processes is the LF 
relativistic spectral
function,
$
{\cal P}^{\tau}_{\sigma'\sigma}(\tilde{\bf \kappa},\epsilon_S,S_{He})
$, with $\tilde{\bf \kappa}$ an intrinsic nucleon momentum and $\epsilon_S$ 
the energy of the two-nucleon
spectator  system. 
The  LF  spectral function is useful also for other studies
(e.g., for nuclear generalized parton distributions, where final states
have to be properly boosted).
With respect to previous attempts
to describe DIS processes off $^3$He in a LF framework (see, e.g.,
the one in Ref. \cite{sauer}), 
here
a special care is devoted to the definition of the
intrinsic light-cone variables as well as to the spin
 degrees of freedom in 
$
{\cal P}^{\tau}_{\sigma'\sigma}(\tilde{\bf \kappa},\epsilon_S,S_{He})~.
$
This PWIA LF spectral function is defined in terms of LF overlaps \cite{PSS} 
between the  ground state of a
polarized $^3$He and
the cartesian product of an interacting state of two nucleons with energy
$\epsilon_S$
and a plane wave for the third
nucleon. Within a reliable approximation \cite{PSS}, it can be
given in terms of the Melosh Rotations, 
$
{{D^{{1 \over 2}} [{\cal R}_M ({\blf \kappa})]}}$,
and the usual {{instant-form spectral function}}
$
{{
{P}^{\tau}_{\sigma'_1\sigma_1}
}}$:
\vskip -1mm
\bq
{ {
{\cal P}^{\tau}_{\sigma'\sigma}({\blf \kappa},\epsilon_{S},S_{He})
}}
\propto  
~\sum_{\sigma_1 \sigma'_{1}} 
{{D^{{1 \over 2}} [{\cal R}_M^\dagger ({\blf
\kappa})]_{\sigma'\sigma'_1}}}~
{{
{P}^{\tau}_{\sigma'_1\sigma_1}({\bf p},\epsilon_{S},S_{He})
}} ~
{{D^{{1 \over 2}} [{\cal R}_M ({\blf \kappa})]_{\sigma_1\sigma}}}  
\eq
\vskip -0.4cm
\begin{table*}
\begin{center}
\label{tab-1}
\begin{tabular}{|c|c|c|c|c|c|}
\hline  &\hspace{-1mm}$proton \, {NR}$\hspace{-2mm}&\hspace{-1mm}$proton \,
{LF}$\hspace{-2mm}&\hspace{-1mm}$neutron \,
{NR}$\hspace{-2mm}&\hspace{-1mm}$neutron \, {LF}$\hspace{-1mm} \\ 
\hline 
$\int^{~}_{~} d \epsilon_S d \tilde{\kappa}\,
~ Tr( {\cal{P}} \sigma_{z})_{\vec{S}_A=
\widehat{z}}$ 
\hspace{3mm} 
& -0.02263 & {{-0.02231}} & 0.87805 & 
{{0.87248}} \\ 
\hline 
\hspace{-7mm} $\int^{~}_{~} d \epsilon_S d \tilde{\kappa}\,
~ Tr( {\cal{P}} \sigma_{y})_{\vec{S}_A=
\widehat{y}}$ \hspace{-2mm}
& -0.02263 & {{-0.02268}} & 0.87805 & {{0.87494}} \\ 
\hline 
\end{tabular}
\end{center} 
\caption{Proton and neutron EPs in PWIA, 
within the non 
relativistic appproach (NR) and within the LF RHD. 
First line: longitudinal EP; second line: 
transverse EP. Preliminary results.}
\end{table*}
\vskip 1mm
Preliminary results are quite encouraging, since, as shown in Tab. 1,
{{LF}} longitudinal and transverse EPs change little from the ones 
obtained within the NR spectral function and weakly differ from
each other. 
Then 
the usual extraction procedure should work well also within 
{{the LF approach}}.
Concerning FSI, we plan to include in our LF description the FSI
between the jet produced from the hadronizing quark
and the two-nucleon spectator system through the GEA introduced in Ref.
\cite{Kaptari}.

\end{document}